\documentclass{elsart}

\usepackage{natbib}

\usepackage{amssymb}

\begin{document}

\begin{frontmatter}

\title{Nonlinear Perpendicular Diffusion in Strong Turbulent Electromagnetic Fields}
\author{Olaf Stawicki}

\address{Unit for Space Physics, School of Physics
North-West University, Potchefstroom Campus, Private Bag X6001,
Potchefstroom 2520, South Africa
}
\ead{fskos@puk.ac.za}
\begin{abstract}
A nonlinear description for perpendicular particle diffusion in strong
electromagnetic fluctuations is developed by using the
fundamental Newton-Lorentz equation. Although not based on the same
approach, the recently presented nonlinear guiding center (NLGC)
theory is recovered as a special case. The approach used here is
rather based on the argument that a well defined particle gyromotion does not
exist in strong fluctuations than on the assumption that
the particle gyrocenter follows magnetic field lines which themselves
separate diffusively. The assumption of a guiding center motion and
the diffusive separation of magnetic field lines is absolutely central
to the NLGC theory. It is argued that the NLGC result should provide
most accurate results for strong fluctuations. Furthermore, as a
direct consequence of the particle equation of motion, it is shown
that particle diffusion in one perpendicular direction is governed by
the fluctuations in the other normal direction. This results
contradicts the NLGC result, where perpendicular diffusion is
triggered by fluctuations in the same direction. This is of particular
interest for anisotropic perpendicular diffusion in a non-axisymmetric
turbulence. Future numerical simulation results for non-axisymmetric
magnetic turbulence and their comparison with the approach presented
here and the NLGC theory have to provide an answer whether
particle diffusion in one perpendicular direction is governed by the
fluctuations in the other normal direction or by fluctuations in the
same direction.
\end{abstract}


\end{frontmatter}

\section{Introduction and Motivation}
\label{sec:intro}
Diffusive particle motion in random fields plays a key role in
space physics, astrophysics and the physics of fusion devices.
Turbulence properties are essential for understanding the
three-dimensional (anisotropic) diffusive particle transport in a
collisionless, turbulent and magnetized plasma such as the solar wind
or the interstellar medium. Of particular interest are the transport
coefficients describing particle diffusion perpendicular to an ambient
magnetic field. 

In spite of its long-standing importance in space and astrophysics,
perpendicular diffusion has been an unsolved puzzle during many
decades and a variety of studies have been carried out to achieve closure 
and to pin it down at a theoretical level. Models have been proposed
for hard-sphere scattering in a magnetized plasma \citep{gleeson1969}
and for related extensions developed on the basis of the Boltzmann
equation \citep{jones1990}, but they seem to be inapplicable to space
plasmas where (electro)magnetic fluctuations trigger particle
scattering \citep[see][]{bm1997}. 

Other models are based on the field line random walk
(FLRW) limit emerging, for slab turbulence geometry, from quasilinear theory
\citep[QLT][]{jokipii1966,jp1969}, which has
been considered and used in subsequent studies 
\citep[e.g.][]{formanetal1974,forman1977,bm1997}. Although FLRW provides for a
physically appealing picture, it has been shown recently that its
applicability is questionable for particle transport in certain
turbulence geometries, particularly those with at least one ignorable
coordinate being the case for slab geometry \citep{jokipiietal1993}. 
Furthermore, numerical simulations, taking into account the magnetic
nature of the turbulence, have shown that FLRW fails to explain perpendicular
transport of low-energy particles
\citep{gj1999,maceetal2000}, since pitch-angle
scattering is neglected in the FLRW limit \citep[e.g.,][]{qinetal2002_apjl}.

It has been argued for some time that diffusion along a background 
magnetic field can reduce perpendicular diffusion to
subdiffusive levels if the turbulence reveals slab geometry only
\citep[see, e.g.,][]{urch1977,kj2000,qinetal2002_grl}. However, when the
turbulent magnetic field has sufficient structure normal to the mean
magnetic field, subdiffusion as seen in pure slab turbulence can be
overcome and diffusion is recovered \citep[see][]{qinetal2002_apjl}. 

Recently, \citet{matthaeusetal2003} proposed the so-called nonlinear
guiding center (NLGC) theory. For their approach, they use the following two key assumptions: First, perpendicular transport is governed by the velocity of
particle gyrocenters that follow magnetic field lines. Second, the
magnetic field lines themselves separate diffusively. Based on these
two assumptions, Matthaeus et al. obtain for the velocity of the
particle gyrocenter in the $x$-direction the expression
\begin{equation} 
v_x =a v_z \frac{\delta B_x}{B_0},
\label{eq:ansatz}
\end{equation}
where $\delta B_x$ is the $x$-component of the turbulent magnetic
field, $B_0$ is the strength of the background magnetic field. The
constant $a$ has to be determined after the fact. Consequently, in the
NLGC theory, particle diffusion in normal direction is solely
triggered by the diffusive separation of the underlying magnetic field
lines. Employing the so-called Taylor-Green-Kubo (TGK) formulation
\citep[see, e.g.,][for more details]{bm1997,matthaeusetal2003},
Matthaeus et al. derive, on the basis of equation (\ref{eq:ansatz}),
the following nonlinear integral equation for the perpendicular
diffusion coefficient:
 \begin{equation} 
\kappa_{xx} =
\frac{a^2v^2}{3B_0^2}\int\limits_{}^{}d^3k\frac{S_{xx}(\bf{k})}{v/\lambda_{zz}+\kappa_{xx}k_{\perp}^2+\kappa_{zz}k_{z}^2+\gamma({\bf
  k})}. 
\label{eq:nlgcres}
\end{equation} 
Here, $\kappa_{zz}=v\lambda_{zz}/3$ is the parallel diffusion
coefficient, $v$ is the particle speed and $\lambda_{zz}$ is the
parallel mean free path. The wavevector components parallel and
perpendicular to the mean magnetic field are given by $k_{z}$ and
$k_{\perp}^2=k_x^2+k_y^2$, respectively, and $S_{xx}$ is the spectral
amplitude of magnetic fluctuations in the $x$-direction. The
decorrelation rate $\gamma({\bf k})$ allows to include dynamical
effects due to the decay of turbulent energy \citep[see, e.g.,][for more details concerning dynamical magnetic
  turbulence]{bieberetal1994}. Since the NLGC theory is based on the
TGK formalism, equation (\ref{eq:nlgcres}) only admits Markovian
diffusion; anomalous transport processes such as subdiffusion and
superdiffusion can not be accommodated by the NLGC model \citep[see
  also][for further discussion]{zanketal2004}.

Considering the particular case of a static ($\gamma({\bf k})=0$)
turbulence, \citet{matthaeusetal2003} use the NLGC approach for weak
($\delta B^2/B_0^2=0.04$) and stronger ($\delta B^2/B_0^2=1.0$)
turbulence and compare the results with numerical simulations. The
numerical simulation results used for the comparison are obtained by
applying a fourth-order Runge-Kutta method with subsequent integration
to the equation of motion of a charged test particle \citep[see][]{qinetal2002_apjl,qinetal2002_grl,matthaeusetal2003}. The finding of the
comparison is that the NLGC theory seems to be consistent with the
numerical simulations for the case $a=1/\sqrt{3}$. Since the NLGC
approach has been fruitful for a better understanding of perpendicular
diffusion, it was recently applied to cosmic ray modulation
and shock acceleration \citep[][]{zanketal2004}.

Although the NLGC approach seems to be in agreement with the numerical
simulation results, several questions arise. Among those are the
following two: (1) Does a well defined particle gyromotion and, therefore, a
gyrocenter actually exists for fluctuations being that strong that the
background magnetic field has no control over the particle motion? 
Wandering in the particle pitch-angle and
gyrophase are then very rapid, and the concept of a gyrocenter
velocity following magnetic field lines is expected to fail for strong
fluctuations. (2) By using the fundamental particle equation of
motion itself, is it actually possible to derive spatial diffusion
coefficients being in agreement with the NLGC result? A closer
inspection of the ansatz (\ref{eq:ansatz}) shows that the NLGC
approach is not developed on the basis of the Newton-Lorentz equation,
which describes particle diffusion in turbulent electromagnetic fields. The
NLGC result, equation (\ref{eq:nlgcres}), is rather based on the
diffusivity of the underlying magnetic field lines. The diffusivity of
magnetic field lines is given by the magnetic field streamline
equation \citep[see, e.g.,][]{jp1969,ruffoloetal2004}, which apparently was
employed to result in equation (\ref{eq:ansatz}).

In view of the two questions given above, a nonlinear description for
perpendicular diffusion in strong electromagnetic fluctuations is
presented which explicitly excludes an ``ordered'' particle
gyromotion. The approach used here is based on the particle equation
of motion and contradicts the above mentioned key assumptions made by
Matthaeus et al. for the NLGC theory. 

 \section{Basic Equations and Formulation}
 \label{sec:deri}
For a random, diffusive particle motion and statistically homogeneous
and stationary fluctuations, $\kappa_{xx}$ is given by the
Taylor-Green-Kubo (TGK) formula \citep{kubo1957} which can be
expressed as
\begin{equation}
\kappa_{xx}=\Re\int\limits_{0}^{\infty}d\xi\langle v_x(0)v_x^{\ast}(\xi)\rangle
\label{eq:tgk1},
\end{equation}
where $v_x$ is the $x$-component of the
particle velocity. An analogous expression is given for the diffusion
coefficient in the $y$-direction, $\kappa_{yy}$. For a large
turbulence coherence time $\xi$, the second-order velocity correlation function
$\langle v_x(0)v_x^{\ast}(\xi)\rangle$ must go to zero, and the integral in
equation (\ref{eq:tgk1}) approaches a constant value for
$t\to\infty$. This is characteristic for Markovian diffusion.

The use of the TGK formula is somewhat critical for situations
where particle transport reveals rather an anomalous diffusion
process than normal Markovian diffusion. In this case, the standard
Fokker-Planck coefficient (anomalous transport law)
$\kappa_{xx}=\langle\Delta x^2\rangle/2t$ applies, where the mean square displacement scales more general as
$\langle\Delta x^2\rangle\propto t^{s}$. Depending on the exponent $s$,
different diffusion processes can be taken into account: $s=1$ for
the diffusive regime, corresponding to a Gaussian random walk of particles;
$s=2$ in the superdiffusive regime, i.e. strictly scatter-free
propagation of the particles \citep[see, e.g.,][]{mccomb1990}, and $s<1$ (e.g. $s=1/2$, as
shown by \citet{qinetal2002_grl} and \citet{kj2000}) in the case of particle
trapping (subdiffusion or compound diffusion). For $s=1$ (normal
Markovian diffusion), the standard Fokker-Planck coefficient is
equivalent to the TGK formula (\ref{eq:tgk1}) for the large $t$
limit.

\subsection{Equations of Motion}
\label{sec:eom}
Generally, the random walk of a particle with mass $m$, charge $q$ and velocity
${\bf v}(t)=\dot{\bf{x}}(t)$ in fluctuating fields is governed by its equation of motion, i.e. the Newton-Lorentz equation 
\begin{equation} 
\dot{\bf p}=\frac{d}{dt}[\gamma m
{\bf v}(t)]=q{\bf E}[{\bf x}(t),t]+\frac{q}{c}{\bf v}(t)\times{\bf B}[{\bf
  x}(t),t],
\label{eq:eom} 
\end{equation} 
where $\gamma$ is the Lorentz factor. The magnetic field ${\bf B}={\bf
  B}_0+\delta{\bf B}$ consists of a uniform, steady background magnetic
  field ${\bf B}_0=B_0{\bf e}_z$ to which is added a broad band
  spectrum of magnetic fluctuations $\delta{\bf B}[{\bf
  x}(t),t]$. Analogously, the electric field ${\bf E}={\bf
  E}_0+\delta{\bf E}$ consists of a vanishing background electric
  field $<{\bf E}>={\bf E}_0={\bf 0}$ with superimposed electric
  fluctuations  $\delta{\bf E}[{\bf x}(t),t]$. Equation (\ref{eq:eom})
  can formally be integrated in time, to obtain for the perpendicular
  velocity components the expressions
\begin{eqnarray} 
v_x(t) & = & 
\frac{q}{\gamma
  mc}\int\limits_{0}^{t}dt^{\prime}\left(c\delta E_x[{\bf
  x}(t^{\prime}),t^{\prime}] +v_y(t^{\prime})\delta B_z[{\bf
  x}(t^{\prime}),t^{\prime}] 
-v_z(t^{\prime})\delta B_y[{\bf
  x}(t^{\prime}),t^{\prime}]\right)
\nonumber
\\
&&+\Omega\int\limits_{0}^{t}dt^{\prime}v_y(t^{\prime}) 
\label{eq:velx}
\\
v_y(t) & = & 
\frac{q}{\gamma
  mc}\int\limits_{0}^{t}dt^{\prime}\left(c\delta E_y[{\bf
  x}(t^{\prime}),t^{\prime}] +v_z(t^{\prime})\delta B_x[{\bf
  x}(t^{\prime}),t^{\prime}] 
-v_x(t^{\prime})\delta B_z[{\bf
  x}(t^{\prime}),t^{\prime}]\right)
\nonumber
\\
&&-\Omega\int\limits_{0}^{t}dt^{\prime}v_x(t^{\prime}), 
\label{eq:vely}
\end{eqnarray} 
where $\Omega=qB_0/\gamma mc$ is the relativistic gyrofrequency. 

The first integrals in equations (\ref{eq:velx}) and (\ref{eq:vely})
describe the random velocity components of the particle in the $x$ and
$y$-direction, respectively, due to the fluctuating fields. They
enter equations (\ref{eq:velx}) and (\ref{eq:vely}) solely by these contributions and represent a
stochastic force acting on the particle and forcing it to deviate from
its unperturbed orbit given by the second integrals. In a uniform and
ordered background magnetic field, each unperturbed orbit is a perfect
helix along which the particle moves at constant speed $v_z$, while
the transverse velocity components $v_x$ and $v_y$ oscillate at
angular frequency $\Omega$ around a magnetic line of force. Now consider
the effects of fluctuations on the unperturbed particle orbit. For
weak ($\delta B^2/B_0^2\ll1$) turbulence conditions, the fluctuations
cause the particle to wander in its pitch-angle and gyrophase. The
force induced by $B_0$ dominates the particle transport and stochastic
contributions resulting from the first integrals are comparably
small. Consequently, a particle ``remembers'' its helical orbit and
performs a diffusive but ordered motion along a preferred direction
defined by ${\bf B}_0$. The deviation from its helical orbit in
perpendicular directions is then marginal, and diffusion is dominated
by parallel scattering. 

In the opposite case of strong ($\delta B^2/B_0^2\geq 1$) turbulence, the
ordered force induced by $B_0$ is small and particle transport is
dominated by the ``chaotic'' contributions given by the first terms in
equations (\ref{eq:velx}) and (\ref{eq:vely}). Then, a particle moves
barely on a perfect trajectory and ``forgets'' its ideal orbit very quickly. 
Decorrelation from the helical orbit then occurs very rapidly, the
wandering in pitch-angle and gyrophase is then purely chaotic, and a
well defined particle gyromotion does not exist. Whether the first 
or second term in equations (\ref{eq:velx})  and (\ref{eq:vely})
dominate depends then solely on the magnitude of the ratio $\delta B^2/B_0^2$. 

The decisive step taken here is to assume strong turbulence conditions,
that the particle has no memory of its formerly helical
trajectory. To do so, the contribution representing the helical orbit
is dropped. This does not imply that ${\bf B}_0={\bf 0}$, it is rather
assumed that the random velocity components (first integral) overwhelm
the unperturbed particle orbit (second integral). Proceeding, the irregular
fields are expressed by Fourier transforms and the first integrals
are, for simplicity and the reason of overview, approximated by their
integrands. This yields
\begin{equation} 
v_x(t) =  
\frac{q\tau_x}{\gamma
  mc}\int\limits_{}^{}d^3ke^{\imath {\bf k}\cdot{\bf x}(t)}\left[c \delta E_x({\bf k},t)+v_y(t)\delta B_z({\bf k},t)
-v_z(t)\delta B_y({\bf k},t)\right],
\label{eq:velxapp}
\end{equation}
\begin{equation} 
v_y(t) =  
\frac{q\tau_y}{\gamma
  mc}\int\limits_{}^{}d^3ke^{\imath {\bf k}\cdot{\bf x}(t)}\left[c \delta E_y({\bf k},t)+v_z(t)\delta B_x({\bf k},t)
-v_x(t)\delta B_z({\bf k},t)\right].
\label{eq:velyapp}
\end{equation}
Here, $\tau_x$ and $\tau_y$ denote timescales related to
the deviation of the particle position from the unperturbed orbit in
the $x$ and $y$-direction, respectively. The timescales $\tau_x$ and
$\tau_y$ might depend on particle rigidity and turbulence properties,
but they are here introduced for the reason of dimensionality. A
future study requires, of course, in any case a detailed treatment of the time integrations. This is far beyond of the scope of
the current paper, since the evaluation is a study on its own. In
particular, the TGK formula (\ref{eq:tgk1}) requires the evaluation of
$v_x$ at time zero, which is a result of the time translation
invariance property usually employed for the derivation of the TGK
formula \citep[see, e.g.,][]{mccomb1990,kj2000}. Obviously, the
substitution of equations (\ref{eq:velx}) and (\ref{eq:vely}) into the
TGK formula would then yield zero for $t=0$. Actually, the approach
used here by employing directly the particle equation of motion
requires the more general Fokker-Planck definition
$\kappa_{xx}=\langle\Delta x^2\rangle/2t$, and this is currently under
investigation. The purpose of this paper is rather to demonstrate that
the same mathematical structure on diffusion coefficients as those of
the NLGC model can be obtained if the Newton-Lorentz equation is used for the
assumption of electromagnetic fluctuations overwhelming effects
related to the background magnetic field. As becomes clear below, the
timescales $\tau_x$ and $\tau_y$ are then related to the ``numerical''
constant $a$ given in equation (\ref{eq:ansatz}). 

 \subsection{Velocity Correlation Functions}
 \label{sec:vel}
The substitution of equations (\ref{eq:velxapp}) and
(\ref{eq:velyapp}) into the TGK formula
(\ref{eq:tgk1}) converts the integrands into a sum of nine individual
contributions involving random particle velocity components and/or
turbulent field components at two different wavevectors. Since the
particle velocity components are random quantities, i.e. $\langle
v_i\rangle=0$, four contributions vanish. Proceeding, it is assumed
that the particle velocity components are uncorrelated with the local
magnetic field vector \citep[see
  also][]{matthaeusetal2003}. Furthermore, supposing that the
probability distribution of the particle displacements and that of the
Eulerian velocity field are statistically independent, which is also
referred to as Corrsin's independence hypothesis
\citep{corrsin1959,matthaeusetal2003,mccomb1990}, one obtains for the
velocity correlation function in $x$-direction the expression
\begin{eqnarray}
\langle v_x(0)v_x^{\ast}(\xi)\rangle & = & \left(\frac{q\tau_x}{\gamma
  mc}\right)^2\int\limits_{}^{}d^3k\int\limits_{}^{}d^3k^{\prime}\left\langle
  e^{-\imath{\bf k}^{\prime}\cdot{\bf
  x}(\xi)}\right\rangle
\Biggl[c^2R_{xx}({\bf k},{\bf k}^{\prime},\xi)
\label{eq:velcorrgen}
\\
&&
+
\langle v_z(0)v_z^{\ast}(\xi)\rangle S_{yy}({\bf k},{\bf k}^{\prime},\xi)
+
\langle v_y(0)v_y^{\ast}(\xi)\rangle S_{zz}({\bf k},{\bf k}^{\prime},\xi)
\nonumber
\\
&&
-
\langle v_y(0)v_z^{\ast}(\xi)\rangle S_{zy}({\bf k},{\bf k}^{\prime},\xi)
-
\langle v_z(0)v_y^{\ast}(\xi)\rangle S_{yz}({\bf k},{\bf k}^{\prime},\xi)
\Biggr]
\nonumber
\end{eqnarray}
and an analogous expression for $\langle v_y(0)v_y^{\ast}(\xi)\rangle$. 
Here, the spectral amplitudes 
\begin{equation}
S_{nm}({\bf k},{\bf
  k}^{\prime},\xi)=\langle\delta B_n({\bf k},0)\delta B_m^{\ast}({\bf
  k}^{\prime},\xi)\rangle
\end{equation}
\begin{equation}
R_{nm}({\bf k},{\bf k}^{\prime},\xi)=\langle\delta
  E_n({\bf k},0)\delta E_m^{\ast}({\bf k}^{\prime},\xi)\rangle
\end{equation}
were introduced, where the subscripts refer to the Cartesian
  coordinates. Furthermore, without loss of generality, the initial
  condition ${\bf x}(0)={\bf 0}$ was used.
For statistically homogeneous turbulence, it is convenient to employ
  the relations $S_{nm}({\bf k},{\bf k}^{\prime},\xi)=\delta({\bf k}-{\bf
  k}^{\prime})S_{nm}({\bf k},\xi)$ and $R_{nm}({\bf k},{\bf
  k}^{\prime},\xi)=\delta({\bf k}-{\bf k}^{\prime})R_{nm}({\bf
  k},\xi)$, and it is assumed that electromagnetic
fluctuations can be represented by a superposition of $N$ individual
plasma wave modes, i.e.
\begin{equation}
\delta {\bf B}({\bf k},t)=\sum\limits_{j=1}^{N}\delta {\bf B}^j({\bf
  k})\exp(-\imath \omega_j t)
\end{equation}
\begin{equation}
\delta {\bf E}({\bf k},t)=\sum\limits_{j=1}^{N}\delta {\bf E}^j({\bf k})\exp(-\imath \omega_j t).
\end{equation}
Here, $\omega_j({\bf k})=\omega_{j,R}({\bf k})+\imath\Gamma_j({\bf
  k})$ is a complex dispersion relation of wave mode $j$, where
  $\omega_{j,R}$ is the real frequency of the mode. The imaginary
  part, $\Gamma_j({\bf k})\leq 0$, represents dissipation of turbulent
  energy due to plasma wave damping. Note that such an ansatz for the
  time variation is usually used in quasilinear theory \citep[see,
  e.g.,][]{gary1993,schlickeiser2002}, where fluctuations are supposed
  to be very weak. Here, however, it is assumed that the same approach
  holds for fluctuations dominating the influence of the background
  magnetic field on particle motion.

Restricting the considerations to transverse fluctuations, i.e. $\delta {\bf E}^j\cdot {\bf k}=0$, and using Faraday's law, the turbulent electric field can easily be expressed by the corresponding magnetic counterparts, yielding
\begin{equation}
\delta E_x^j=\frac{\omega_j}{ck^2}\left(\delta B_y^jk_{\|}-\delta B_{\|}^jk_y\right)
\,\,\,;\hspace*{0.1cm}
\delta E_y^j=\frac{\omega_j}{ck^2}\left(\delta B_{\|}^jk_x-\delta B_x^jk_{\|}\right)
\label{eq:fara}
\end{equation}
and, subsequently, the relation
\begin{equation}
R_{xx}=\frac{|\omega_j|^2}{c^2k^4}\left[S_{yy}^jk_{\|}^2+S_{zz}^jk_{y}^2-k_{\|}k_y\left(S_{zy}^j+S_{yz}^j\right)\right].
\end{equation}
For simplicity, it is assumed that $S_{zy}^j({\bf
  k})=S_{yz}^j({\bf k})=0$ for following calculations. In what follows, one arrives at
\begin{eqnarray}
\langle v_x(0)v_x^{\ast}(\xi)\rangle & = & \left(\frac{q\tau_x}{\gamma
  mc}\right)^2\sum\limits_{j=1}^{N}\int\limits_{}^{}d^3k\left\langle e^{-\imath{\bf k}\cdot{\bf
  x}(\xi)}\right\rangle e^{\imath\omega_{j,R}({\bf k})\xi+\Gamma_j({\bf k})\xi}
\label{eq:velcorrsim}
\\
&&\times
\Biggl[
\frac{|\omega_j|^2}{k^4}\left(k_{\|}^2S_{yy}^j({\bf k})+k_{y}^2S_{zz}^j({\bf k})\right)
\nonumber
\\
&&
+
\langle v_z(0)v_z^{\ast}(\xi)\rangle S_{yy}^j({\bf k})
+
\langle v_y(0)v_y^{\ast}(\xi)\rangle S_{zz}^j({\bf k})
\Biggr]
\nonumber
\end{eqnarray}
and obtains an analogous expression for $\langle v_y(0)v_y^{\ast}(\xi)\rangle$.

 \subsection{Nonlinear Diffusion Coefficients}
 \label{sec:diffco}
One can now proceed to derive expressions for
$\kappa_{xx}$ as well as $\kappa_{yy}$. To do so, the time-integration in equation
(\ref{eq:tgk1}) has to be performed, and further progress requires the
variations of the individual contributions in the coherence time $\xi$.
First, the velocity correlation functions in
equation (\ref{eq:velcorrsim}) can be written as \citep[see also][]{matthaeusetal2003}
\begin{equation}
\langle v_{n}(0)v_{n}^{\ast}(\xi)\rangle=(v^2/3)\exp(-v\xi/\lambda_{nn}),
\end{equation}
where $\kappa_{nn}=v\lambda_{nn}/3$ is used, being in agreement with
the TGK definition (\ref{eq:tgk1}). Here, $\lambda_{nn}$ is
the mean free path in the direction of one of the Cartesian
coordinates. Second, the ensemble-averaged exponential expression can be expressed as
\begin{equation}
\left\langle\exp\left[-\imath{\bf k}\cdot{\bf
	x}(\xi)\right]\right\rangle=\int\limits_{-\infty}^{\infty}dx\int\limits_{-\infty}^{\infty}dy\int\limits_{-\infty}^{\infty}dz
	P(x,y,z)e^{-\imath
	k_x x-\imath k_y y-\imath k_z z},
\label{eq:prob}
\end{equation}
where the distribution function $P(x,y,z)$ represents displacements
of the particle position from the unperturbed orbit in Cartesian space
due to the decorrelation of the trajectory. The decisive step is to
assume a Gaussian distribution, i.e. 
\begin{equation}
P(x,y,z)=\frac{1}{(2\pi)^{3/2}\sigma_x\sigma_y\sigma_z}\exp\left[-\frac{x^2}{2\sigma_x^2}-\frac{y^2}{2\sigma_y^2}-\frac{z^2}{2\sigma_z^2}\right],
\label{eq:gaussian}
\end{equation}
where the variances $\sigma_x^2=\langle\Delta x^2\rangle$,
$\sigma_y^2=\langle\Delta y^2\rangle$ and $\sigma_z^2=\langle\Delta
z^2\rangle$ are assumed to be diffusive in the sense that
$\langle\Delta x^2\rangle=2\kappa_{xx}\xi$, $\langle\Delta
y^2\rangle=2\kappa_{yy}\xi$ and $\langle\Delta z^2\rangle=2\kappa_{zz}\xi$.
Upon substituting the Gaussian distribution function
(\ref{eq:gaussian}) into equation (\ref{eq:prob}), one arrives, after elementary integrations, at
\begin{equation}
\left\langle\exp\left[-\imath{\bf k}\cdot{\bf
	x}(\xi)\right]\right\rangle=\exp[-(\kappa_{xx}k_{x}^2+\kappa_{yy}k_{y}^2+\kappa_{zz}k_{z}^2)\xi].
\label{eq:prob2}
\end{equation}
Using these two entries, the $\xi$-integration in
equation (\ref{eq:tgk1}) is elementary. Upon extending in
equation (\ref{eq:velcorrsim}) the factor in front of the
wavevector integral by $B_0$ (not assumed to be zero), one finally
obtains for $\kappa_{xx}$ and $\kappa_{yy}$ the following expressions:
\begin{eqnarray} 
\kappa_{xx} & = &
\frac{a_x^2v^2}{3B_0^2}\sum\limits_{j=1}^{N}\int\limits_{}^{}d^3k
\Biggl[\frac{3|\omega_j|^2}{v^2k^4}\left(k_{\|}^2S_{yy}^j({\bf k})+k_{y}^2S_{zz}^j({\bf k})\right)\frac{A}{A^2+\omega_{j,R}^2}
\label{eq:kappax}
\\
&&
+
S_{yy}^j({\bf
	k})\frac{A+v/\lambda_{zz}}{(A+v/\lambda_{zz})^2+\omega_{j,R}^2}
+
S_{zz}^j({\bf
	k})\frac{A+v/\lambda_{yy}}{(A+v/\lambda_{yy})^2+\omega_{j,R}^2}\Biggr],
\nonumber
\\[0.3cm]
\kappa_{yy} & = &
\frac{a_y^2v^2}{3B_0^2}\sum\limits_{j=1}^{N}\int\limits_{}^{}d^3k
\Biggl[\frac{3|\omega_j|^2}{v^2k^4}\left(k_{\|}^2S_{xx}^j({\bf k})+k_{x}^2S_{zz}^j({\bf k})\right)\frac{A}{A^2+\omega_{j,R}^2}
\label{eq:kappay}
\\
&&
+
S_{xx}^j({\bf
	k})\frac{A+v/\lambda_{zz}}{(A+v/\lambda_{zz})^2+\omega_{j,R}^2}
+
S_{zz}^j({\bf
	k})\frac{A+v/\lambda_{xx}}{(A+v/\lambda_{xx})^2+\omega_{j,R}^2}\Biggr].
\nonumber
\end{eqnarray} 
Here, $A=\kappa_{xx}k_{x}^2+\kappa_{yy}k_{y}^2+\kappa_{zz}k_{z}^2-\Gamma_j({\bf k})$ and
$a_x=\tau_x\Omega$ as well as $a_y=\tau_y\Omega$ are introduced. Equations (\ref{eq:kappax}) and (\ref{eq:kappay}), the
main results of this paper, form a set of two nonlinear and coupled equations 
for the two diffusion coefficients in the two distinguished perpendicular
directions. Each diffusion coefficient depends not only on
$\kappa_{zz}$, but also on the perpendicular transport parameter
describing (diffusive) particle transport in the other normal direction. 

An interpretation of the individual contributions can be given by the ``right-hand'' rule known from
electrodynamics. The scattering of a particle in a certain Cartesian
direction can be considered as the result of a stochastic force acting
on the particle and forcing it to move diffusively in this
direction. For instance, consider the contributions appearing
in $\kappa_{xx}$. According to the Lorentz-force
term in equation (\ref{eq:eom}), a ``chaotic'' force in $x$-direction requires a
random particle speed along the $z$-axis. In the second contribution in
equation (\ref{eq:kappax}), we see that there is a stochastic velocity
component $v_z$ present, through $\lambda_{zz}$. This random speed,
together with the irregular magnetic fluctuation in the
$y$-direction, via $S_{yy}^j$, results in the diffusion in the
$x$-direction. Particle diffusion in one perpendicular direction is, therefore,
governed by the fluctuations in the other normal direction. The
third contribution results from the turbulent field component aligned
along the weak background magnetic field. According to the Lorentz-force
term, a random particle speed in $y$-direction
is then required or, equivalently, the mean free path
$\lambda_{yy}$. Finally, the first term in equation (\ref{eq:kappax}) results
from Faraday's law by which the fluctuating electric field in
$x$-direction is expressed by the corresponding magnetic field components.

It is instructive to compare the general diffusion coefficient
(\ref{eq:kappax}) with the NLGC result by \citet{matthaeusetal2003} in
equation (\ref{eq:nlgcres}). For their derivation, they assume that the
fluctuations are purely magnetic. The limit of vanishing random electric fields can be achieved by choosing $\omega_{j,R}({\bf k})=0$ in
equation (\ref{eq:kappax}), initially derived for the
plasma wave viewpoint. The first term in equation (\ref{eq:kappax})
results from Faraday's law and, therefore, would not occur for purely
magnetic fluctuations. Consequently, it is dropped for the
comparison. Since the concept of a superposition of individual wave
modes does not apply, $N=1$, and the corresponding
$j=1$-sub(super)scripts are omitted. For the dynamical behavior
of purely magnetic fluctuations, Matthaeus et al. introduced the
quantity $\gamma({\bf k})$. To take this into account for the
comparison, the plasma wave dissipation rate $\Gamma_j({\bf k})\leq 0$ has
to be replaced by $-\gamma({\bf k})$. For purely magnetic
fluctuations, one then obtains the general relations 
\begin{equation} 
\kappa_{xx}=\frac{a_x^2v^2}{3B_0^2}\int\limits_{}^{}d^3k\left[\frac{S_{yy}(\bf{k})}{B+v/\lambda_{zz}+\gamma({\bf
 k})}+\frac{S_{zz}(\bf{k})}{B+v/\lambda_{yy}+\gamma({\bf
 k})}\right]
\label{eq:kappax2}
\end{equation} 
\begin{equation} 
\kappa_{yy} = \frac{a_y^2v^2}{3B_0^2}\int\limits_{}^{}d^3k\left[\frac{S_{xx}(\bf{k})}{B+v/\lambda_{zz}+\gamma({\bf
 k})}+\frac{S_{zz}(\bf{k})}{B+v/\lambda_{xx}+\gamma({\bf
 k})}\right], 
\label{eq:kappay2}
\end{equation}
where $B=\kappa_{xx}k_{x}^2+\kappa_{yy}k_{y}^2+\kappa_{zz}k_{z}^2$.
For their approach, Matthaeus et
al. neglect fluctuations along the background magnetic field, implying
here $S_{zz}=0$. Their assumption of an axisymmetric turbulence can
be achieved by using $\tau_x=\tau_y$, $S_{xx}=S_{yy}$ and
$\kappa_{xx}=\kappa_{yy}$.  The diffusion coefficient
(\ref{eq:kappax2}) then reduces to the NLGC result. The only difference
is the term in front of the integral, and a comparison leads to the
important result that $a=\tau_x\Omega$. Matthaeus et al. argue that
their factor $a$ is related to the gyrocenter velocity. Numerically, they found
$a=1/\sqrt{3}$, implying here $\tau_x\Omega=1/\sqrt{3}$. But this is
characteristic of strong turbulence, where an ``ordered'' and well
defined gyromotion of charged particles is not expected to be present
anymore. Equations (\ref{eq:kappax}) and (\ref{eq:kappay}) and their
simplified versions (\ref{eq:kappax2}) and (\ref{eq:kappay2}),
respectively, are, therefore, referred to as the strong nonlinear
(SNL) theory. The considerations presented above and the formal agreement of
the simplified SNL equations with the NLGC result indicate that the
latter will probably provide for most accurate results if strong
fluctuations are assumed.

 \section{Summary and Conclusions}
 \label{sec:sum}
The focus of this paper is nonlinear perpendicular diffusion of
charged particles in irregular electromagnetic fields. Using the
fundamental Newton-Lorentz equation for the random motion of a test
particle and assuming that a well defined particle gyromotion does not
exist in strong turbulence, nonlinear integral equations are derived
representing particle diffusion in the two distinct perpendicular
directions. The two new nonlinear diffusion coefficients are valid for a
plasma wave dispersion relation, i.e. real frequency of a plasma wave
mode and its decay due to dissipation, depending arbitrarily on
wavevector. Both transport parameters are not only coupled with the
diffusion coefficient for parallel particle transport, but they are
also coupled among one another. The new set of nonlinear perpendicular
diffusion coefficients is referred to as the strong nonlinear (SNL)
theory, and the NLGC theory by \citet{matthaeusetal2003} is recovered for
the case of purely magnetic fluctuations. Although the SNL theory
generalizes the NLGC approach and, furthermore, is based on another
physically appealing approach, namely the Newton-Lorentz equation, a
complete theory requires a prescription for determining the input
parameters $a_x=\tau_x\Omega$ and $a_y=\tau_y\Omega$ or, better, the
detailed evaluation of the time-integrations in equations (\ref{eq:velx}) and
(\ref{eq:vely}). Of particular interest is the observation that
particle diffusion in one perpendicular direction is governed by the
fluctuations in the other normal direction, e.g., $\kappa_{xx}\propto
S_{yy}$. This is a direct consequence of the Lorentz-term appearing in
the Newton-Lorentz equation (\ref{eq:eom}). On the other hand, the
NLGC result, equation (\ref{eq:nlgcres}), predicts $\kappa_{xx}\propto
S_{xx}$. This difference might be of interest for perpendicular
diffusion in non-axisymmetric turbulence, since $S_{xx}$ and $S_{yy}$
are then not equal \citep[e.g.,][]{ruffoloetal2001}. Future numerical
simulation results for non-axisymmetric magnetic turbulence and their
comparison with equations (\ref{eq:kappax2}) and (\ref{eq:kappay2})
as well as the NLGC theory should then provide for an answer whether
particle diffusion in one perpendicular direction is governed by the
fluctuations in the other normal direction or by fluctuations in the
same direction.


\begin{thebibliography}{24}
\bibitem[{Bieber and Matthaeus(1997)}]{bm1997}
Bieber, J.~W., Matthaeus, W.~H., 1997. Perpendicular diffusion and drift at
  intermediate cosmic-ray energies. Astrophys. J. 485~(2), 655--659.

\bibitem[{Bieber et~al.(1994)Bieber, Matthaeus, Smith, Wanner, Kallenrode, and
  Wibberenz}]{bieberetal1994}
Bieber, J.~W., Matthaeus, W.~H., Smith, C.~W., Wanner, W., Kallenrode, M.-B.,
  Wibberenz, G., 1994. Proton and electron mean free paths: {T}he {P}almer
  consensus revisited. Astrophys. J. 420~(1), 294--306.

\bibitem[{Corrsin(1959)}]{corrsin1959}
Corrsin, S., 1959. Atmospheric diffusion and air pollution. In: Frenkiel, F.,
  Sheppard, P. (Eds.), Advances in Geophys. Vol.~6. New York: Academic Press,
  p. 161.

\bibitem[{Forman(1977)}]{forman1977}
Forman, M.~A., 1977. The velocity correlation function in cosmic ray diffusion
  theory. Astrophys. Space Sci. 49~(1), 83--97.

\bibitem[{Forman et~al.(1974)Forman, Jokipii, and Owens}]{formanetal1974}
Forman, M.~A., Jokipii, J.~R., Owens, A.~J., 1974. Cosmic-ray streaming
  perpendicular to the mean magnetic field. Astrophys. J. 192~(2), 535--540.

\bibitem[{Gary(1993)}]{gary1993}
Gary, S.~P., 1993. Theory of {S}pace {P}lasma {M}icroinstabilities. Cambridge
  University Press, Cambridge.

\bibitem[{Giacalone and Jokipii(1999)}]{gj1999}
Giacalone, J., Jokipii, J.~R., 1999. The transport of cosmic rays across a
  turbulent magnetic field. Astrophys. J. 520~(1), 204--214.

\bibitem[{Gleeson(1969)}]{gleeson1969}
Gleeson, L.~J., 1969. The equations describing the cosmic-ray gas in the
  interplanetary region. Planet. Space Sci. 17, 31--47.

\bibitem[{Jokipii(1966)}]{jokipii1966}
Jokipii, J.~R., 1966. Cosmic-ray propagation. {I}. {C}harged particles in a
  random magnetic field. Astrophys. J. 146~(2), 480--487.

\bibitem[{Jokipii et~al.(1993)Jokipii, K\'ota, and Giacalone}]{jokipiietal1993}
Jokipii, J.~R., K\'ota, J., Giacalone, J., 1993. Perpendicular transport in 1-
  and 2-dimensional shock simulations. Geophys. Res. Lett. 20~(17), 1759--1761.

\bibitem[{Jokipii and Parker(1969)}]{jp1969}
Jokipii, J.~R., Parker, E.~N., 1969. Stochastic aspects of magnetic lines of
  force with application to cosmic-ray propagation. Astrophys. J. 155~(3),
  777--798.

\bibitem[{Jones(1990)}]{jones1990}
Jones, F.~C., 1990. The generalized diffusion-convection equation. Astrophys.
  J. 361, 162--172.

\bibitem[{K{\'o}ta and Jokipii(2000)}]{kj2000}
K{\'o}ta, J., Jokipii, J.~R., 2000. Velocity correlation and the spatial
  diffusion coefficients of cosmic rays: compound diffusion. Astrophys. J.
  531~(2), 1067--1070.

\bibitem[{Kubo(1957)}]{kubo1957}
Kubo, R., 1957. Statistical-mechanical theory of irreversible processes. i.
  general theory and simple applications to magnetic and conduction problems.
  J. Phys. Soc. Jpn. 12, 570--586.

\bibitem[{Mace et~al.(2000)Mace, Matthaeus, and Bieber}]{maceetal2000}
Mace, R.~L., Matthaeus, W.~H., Bieber, J.~W., 2000. Numerical investigation of
  perpendicular diffusion of charged test particles in weak magnetostatic slab
  turbulence. Astrophys. J. 538~(1), 192--202.

\bibitem[{Matthaeus et~al.(2003)Matthaeus, Qin, Bieber, and
  Zank}]{matthaeusetal2003}
Matthaeus, W.~H., Qin, G., Bieber, J.~W., Zank, G.~P., 2003. Nonlinear
  collisionless perpendicular diffusion of charged particles. Astrophys. J.
  Lett. 590~(1), L53--L56.

\bibitem[{McComb(1990)}]{mccomb1990}
McComb, W.~D., 1990. The Physics of Fluid Turbulence. Clarendon Press, Oxford.

\bibitem[{Qin et~al.(2002{\natexlab{a}})Qin, Matthaeus, and
  Bieber}]{qinetal2002_apjl}
Qin, G., Matthaeus, W.~H., Bieber, J.~W., 2002{\natexlab{a}}. Perpendicular
  transport of charged particles in composite model turbulence: Recovery of
  diffusion. Astrophys. J. Lett. 578, L117--L120.

\bibitem[{Qin et~al.(2002{\natexlab{b}})Qin, Matthaeus, and
  Bieber}]{qinetal2002_grl}
Qin, G., Matthaeus, W.~H., Bieber, J.~W., 2002{\natexlab{b}}. Subdiffusive
  transport of charged particles perpendicular to the large scale magnetic
  field. Geophys. Res. Lett. 29~(4), 10.1029/2001{GL}014035.

\bibitem[{Ruffolo et~al.(2001)Ruffolo, Chuychai, and
  Matthaeus}]{ruffoloetal2001}
Ruffolo, D., Chuychai, P., Matthaeus, W.~H., 2001. Field line random walk for
  non-axisymmetric magnetic fluctuations. In: Proc. Int. Conf. Cosmic Ray 27th,
  \textnormal{Hamburg, Germany}. pp. 3729--3731.

\bibitem[{Ruffolo et~al.(2004)Ruffolo, Matthaeus, and
  Chuychai}]{ruffoloetal2004}
Ruffolo, D., Matthaeus, W.~H., Chuychai, P., 2004. Separation of magnetic field
  lines in two-dimensional turbulence. Astrophys. J. 614, 420--434.

\bibitem[{Schlickeiser(2002)}]{schlickeiser2002}
Schlickeiser, R., 2002. Cosmic Ray Astrophysics. Springer, Berlin.

\bibitem[{Urch(1977)}]{urch1977}
Urch, I.~H., 1977. Charged particle transport in turbulent magnetic fields:
  {T}he perpendicular diffusion coefficient. Astrophys. Space Sci. 46,
  389--406.

\bibitem[{Zank et~al.(2004)Zank, Gang, Florinski, Matthaeus, Webb, and
  le~Roux}]{zanketal2004}
Zank, G.~P., Gang, L., Florinski, V., Matthaeus, W.~H., Webb, G.~M., le~Roux,
  J.~A., 2004. Perpendicular diffusion coefficient for charged particles of
  arbitrary energy. J. Geophys. Res. 109~(A04107), doi:10.1029/2003JA010301.

\end{thebibliography}

\end{document}